\documentclass[12pt]{iopart}
\usepackage{cite}
\usepackage{graphicx}
\usepackage{float}
\usepackage{longtable}

\begin{document}

\title[In-situ conductometer for characterizing high-temperature and high-pressure brines]{An in-situ conductometric apparatus for physicochemical characterization of solutions and in-line monitoring of separation processes at elevated temperatures and pressures}

\author{Tae Jun Yoon, Jacob D Riglin, Prashant Sharan, Robert P Currier, Katie A Maerzke, and Alp T Findikoglu}
\address{Los Alamos National Laboratory, Los Alamos, NM 87545, the United States}
\ead{tyoon@lanl.gov}
\vspace{10pt}
\begin{indented}
\item[]September 2021
\end{indented}

\begin{abstract}
	Specific conductance and frequency-dependent resistance (impedance) data are widely utilized for understanding the physicochemical characteristics of aqueous and non-aqueous fluids and for evaluating the performance of chemical processes. However, the implementation of such an \textit{in-situ} probe in high-temperature and high-pressure environments is not trivial. This work provides a description of both the hardware and software associated with implementing a parallel-type \textit{in-situ} electrochemical sensor. The sensor can be used for in-line monitoring of thermal desalination processes and for impedance measurements in fluids at high temperature and pressure. A comparison between the experimental measurements on the specific conductance in aqueous sodium chloride solutions and the conductance model demonstrate that the methodology yields reasonable agreement with both the model and literature data. A combination of hardware components, a software-based correction for experimental artifacts, and computational fluid dynamics (CFD) calculations used in this work provide a sound basis for implementing such \textit{in-situ} electrochemical sensors to measure frequency-dependent resistance spectra.
\end{abstract}

%
%
%
\maketitle
%
%

\section{Introduction}
The response of a material to an external electromagnetic field is widely utilized for both scientific and industrial purposes. In the chemical industries, the specific conductance, which is defined as the ability to conduct electricity under direct current (DC), is routinely used as a quick and reliable method to evaluate the process performance. For instance, it has been used to monitor the performance of desalination processes, corrosion, fuel cell health \cite{halvorsen2020electrochemical}, alcohol fermentation \cite{li2019analysis}, sterilization \cite{oberlander2018spore}, and food product control \cite{dorai2001influence}. The electrical properties, including the dielectric constant and equivalent conductance, are utilized to understand dissolution and/or association behavior of solutes, solution heterogeneities, and kinetic properties in solution (e.g., mutual diffusion).

Both \textit{in-situ} and \textit{ex-situ} conductometric techniques can be utilized to evaluate process performance or monitor the phenomenon of interest within a chemical system. Compared to \textit{in-situ} spectroscopic techniques, \textit{ex-situ} conductometry can be implemented easily. Numerous commercial conductometers are available for measuring the specific conductance at a specific frequency (typically $\mathcal{O}(10^3)$ Hz) \cite{bevster2006modern}. Since the physicochemical characteristics of many aqueous systems are known under ambient conditions, data interpretation is straightforward in such cases. For more complex systems consisting of an unknown chemical species, selective electrodes are often required.

Compared to \textit{ex-situ} analyses, an \textit{in-situ} electrochemical impedance spectroscopy (\textit{in-situ} EIS) offers distinct advantages. Above all, complex resistance (impedance) spectra provide a wealth of information/insight. The low-frequency response can be utilized for rapid evaluation of material corrosion,\cite{ningshen2009electrochemical,rodriguez2017accelerated} diffusion characteristics of solutes,\cite{nguyen2018determination} and structural characteristics of the double layer \cite{ishai2013electrode}. As the excitation frequency increases, the real part of the impedance decreases, whereas its imaginary part increases. This change can be used to estimate electrical properties of the bulk phase, including the resistance-to-capacitance (RC) characteristic time $\tau_\mathrm{RC}$, dielectric constant $\epsilon_\mathrm{r}$, and specific conductance $\kappa$. Based on the theoretical or semi-empirical models for electrolyte solutions, these physical properties can then be used to estimate thermodynamic and kinetic properties (e.g., hydration number, diffusivity, and association constant) at process operating conditions. In addition to these properties, another noteworthy advantage of \textit{in-situ} EIS is its ability for in-line monitoring of reaction and/or separation. In separation processes such as the desalination processes and liquid-liquid extraction processes, it is often desirable to achieve a selective precipitation to recover valuable or harmful materials. Differences in physico-chemical properties between electrolytes and their complexation behaviors in solution are important to understand for designing such a separation process An \textit{in-situ} electrochemical sensor can provide such information in a variety of chemical or biochemical processes \cite{guth2009recent,o2015non,margueres2013preliminary,alig2010monitoring}.

Notwithstanding these advantages, it is technically challenging to implement an \textit{in-situ} electrochemical sensor in high temperature and pressure environments. Since the sensor must typically be positioned in a high-temperature and high-pressure vessel, the data can be easily corrupted by a variety of measurement-related artifacts such as low-quality signal due to the electrical leakage (or short resisdence) and electrode contamination and/or degradation. Considering these challenges, this work reports on the implementation of an \textit{in-situ} conductometric technique for use in a continuous high-temperature and high-pressure thermal desalination unit. The goal is reliable measurement of impedance spectra in aqueous brines at high temperature and pressure. 
\section{Methods}
\subsection{Materials}
Distilled water (Kroger\textsuperscript{\tiny\textregistered}) was purchased locally. The specific conductance of the distilled water at ambient conditions was always measured as $<1\ \mathrm{{\mu}S/cm}$ by a commercial electrode at 293 K (EW-19601-03, Traceable\textsuperscript{\tiny\textregistered}). Sodium chloride (NaCl, Sigma Aldrich, $\geq99.9\ \%$) was purchased from Sigma Aldrich. In preparing solutions, sodium chloride and water were weighed by an electronic balance (MS204S/03, Mettler Toledo) prior to mixing. After mixing, a calibrated sodium selective electrode (LAQUAtwin Na-11 ion meter, Horiba\textsuperscript{\tiny\textregistered}) was used to determine the NaCl concentration in a solution. Citric acid ($\geq99.5\ \%$, anhydrous, free-flowing, Redi-Dri\textsuperscript{TM}, ACS reagent) used for surface passivation was supplied from Sigma Aldrich. Platinizing solution (YSI 3140 Platinizing Solution, YSI) was obtained from Yellow Spring Instruments. All substances were used as received without additional purification.
\subsection{Continuous flow apparatus}
\begin{figure*}
	\includegraphics[width=\textwidth]{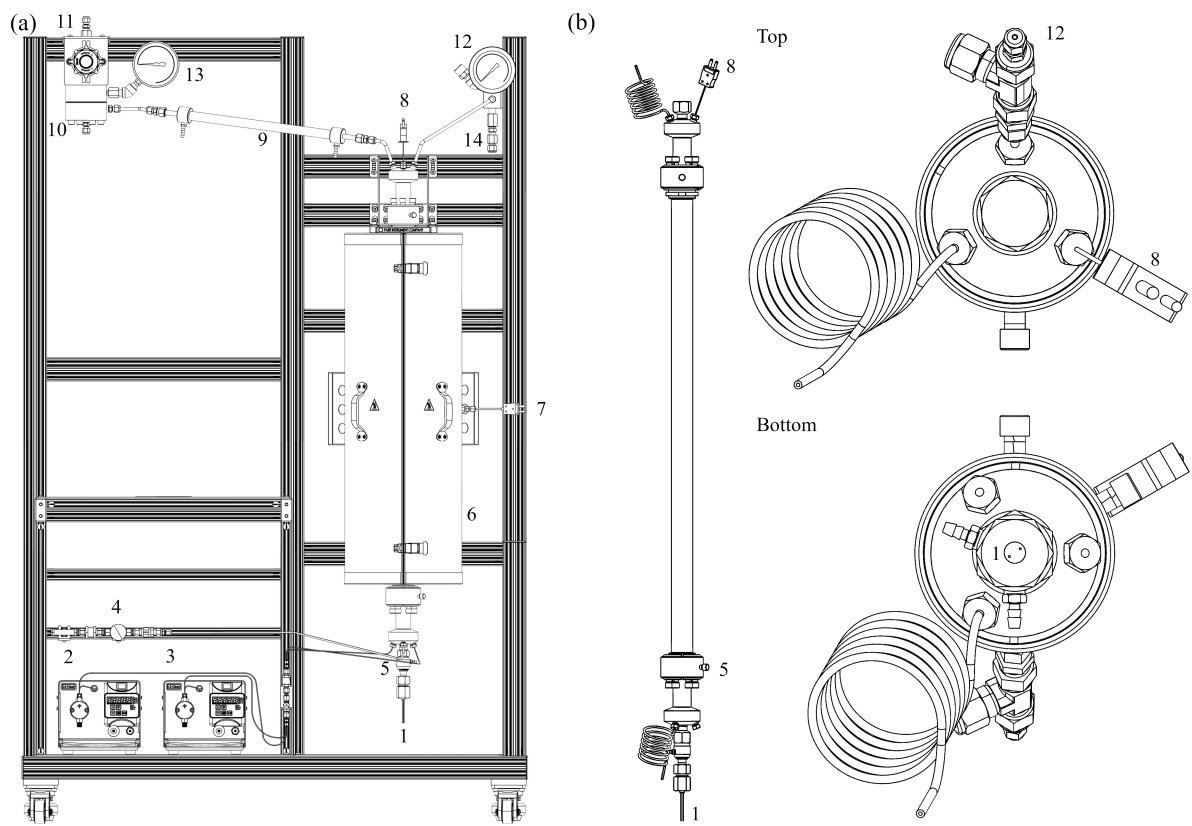}
	\caption{Schematics of (a) the continuous supercritical water desalination unit and (b) the continuous flow cell equipped with electrodes for the measurement of the specific conductance: 1, electrical feedthrough; 2 and 3, LS-class HPLC pumps; 4, purge line; 5, cooler; 6, split-tube type heater; 7, outside thermocouple; 8, inside thermocouple; 9, single-pass heat exchanger; 10, back pressure regulator; 11, forward pressure regulator (dome); 12, vessel pressure indicator; 13, dome pressure indicator; 14, pressure relief device.}
	\label{fig1:apparatus}
\end{figure*}

Figure \ref{fig1:apparatus} shows a schematic diagram of the continuous thermal desalination unit designed and built for this work. This unit is similar to the continuous reactor system offered by Parr (Model 5400, Parr Instrument Company\textsuperscript{\tiny\textregistered}). Two high-pressure pumps (LS-Class Pump 903034 REV L, Teledyne SSI) were installed to supply the feed(s) into a vertical cell of length 0.822 m. The entire volume of the flow unit was approximately 450 mL. The cell temperature was controlled by a split-tube type electrical heater (Model A3653HC20EE, ThermCraft\textsuperscript{\tiny\textregistered}) in conjunction with a temperature controller (4848 Reactor Controller, Parr Instrument Company\textsuperscript{\tiny\textregistered}). Two thermocouples were installed to measure and control the operating temperature. One thermocouple of length 0.406 m was inserted into the system directly contacting the working fluid. The other measured the surface temperature of the cell. The outputs from these thermocouples were delivered to the temperature controller so that the heating power could be automatically adjusted to force the inner thermocouple temperature to the specified set point.
After passing through the cell, the process fluid was cooled by a single-pass heat exchanger. The temperature of the single-pass heat exchanger was maintained at 283 K by connecting it to a commercial chiller (RTE-111, Neslab\textsuperscript{\tiny\textregistered}). Water was used as a coolant. The system pressure was maintained by a Teflon\textsuperscript{\tiny\textregistered} diaphragm-type pressure regulator (A93VB, Parr Instrument Company\textsuperscript{\tiny\textregistered}). The regulator dome was controlled by supplying argon via a forward pressure regulator (Model 44-1166-24, Tescom\textsuperscript{\tiny\textregistered}) connected to a 413 bar (6,000 psi) argon cylinder ($\geq99.999\ \%$, Matheson Tri-Gas\textsuperscript{\tiny\textregistered}). The system pressure was monitored at two points, at the top of the cell and the dome of the back pressure regulator. The operating temperature and pressure limits of the system were 773 K and 345 bar (5,000 psi), respectively. 

All parts of the system exposed to high-temperature brines were made of Hastelloy\textsuperscript{\tiny\textregistered} C276. which has superior corrosion resistance at elevated temperatures and pressures, compared to stainless steel and Inconel alloys \cite{tang2015corrosion}. Nevertheless, this alloy did show some evidence of corrosion when exposed to highly concentrated brines at elevated temperatures. To enhance corrosion resistance, chemical passivation of all parts exposed to the brine were performed to form protective oxide layers. Following Yasensky et al. \cite{yasensky2009citric}, citric acid solution (4 \textit{wt}\%) was continuously flowed into the reactor at ambient pressure for 2 hours while maintaining the cell temperature at 353 K.
\subsection{Preparation and installation of the \textit{in-situ} probe}
Since the \textit{in-situ} measurement electrodes had to be directly inserted into a high-temperature and high-pressure environment, we prepared and implemented the probe with the following considerations. First, the electrode material had to be robust in the corrosive environment. The formation of an oxide scale on the electrode surface can introduce an unwanted artifact to the measurements. Thus, we prepared parallel plate platinized platinum electrodes (Figure 2). The physical dimension of an electrode plate was 1.5 cm by width and height. Platinization was performed by first cleaning the electrode surface with diluted \textit{aqua regia} (50 \% by volume). The cleaned electrodes were then dipped into a platinizing solution, and coated with platinum black by applying a current density of 30 mA/cm\textsuperscript{2} for ten minutes \cite{feltham1971platinized}. The deposition current was controlled by connecting the electrodes to a commercial power supplier (Power station 1010, Plating Electronic).

\begin{figure}
	\begin{center}
	\includegraphics[width=0.45\textwidth]{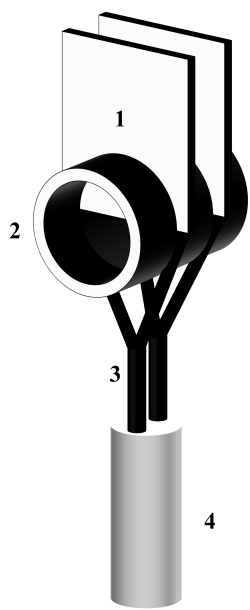}
	\caption{Schematics of the \textit{in-situ} conductometric sensor: 1, platinized platinum electrode; 2, aluminum oxide support (separator); 3, titanium grade 5 (Ti-6Al-4V) wires (outer diameter 1.628 mm); 4, two-bore alumina rod.}
	\label{fig2: electrode}
	\end{center}
\end{figure}

The platinized electrodes were then welded (Model 1-163-01-03, Unitek\textsuperscript{\tiny\textregistered}) to a pair of titanium alloy wires (14 AWG Ti-6Al-4V, Nexmetal Corporation). Ti-6Al-4V alloy was selected since it usually demonstrates good corrosion resistance in high-temperature aqueous environments \cite{tang2015corrosion}. The wires were oxidized for 8 hours at 600 $^\circ$C in air in order to reduce their corrosion rate in a wet environment \cite{velten2002preparation}. After the oxidation passivation step, the wire surface was covered with an apparently uniform oxide layer. The wires were then inserted into the cell via an electrical feedthrough (PL-14-A-2, Conax Technologies). The electrical feedthrough consisted of a cap, gland body, Teflon\textsuperscript{\tiny\textregistered} sealant, and ceramic insulators for the insulation of the wires from the cell surface. Since the Teflon\textsuperscript{\tiny\textregistered} sealant (6554-130, Conax Technologies) can be vulnerable to a high-temperature environment, the gland was cooled by installing a cooling section between the electrical feedthrough and the cell and flowing coolant at 283 K (Figure 1b). To immobilize the location of the electrode and to protect the wire surface, the wires were covered with a two-bore alumina rod (AL-T2-N156-N04-30, Advalue Technology), and the electrodes were inserted into slits made in a hollow-aluminum oxide rod so that the distance between the electrodes was kept fixed (Figure \ref{fig2: electrode}).
Since electrodes prepared in the manner described above could still degrade when exposed to a high-temperature environment \cite{zimmerman1995new}, they were exposed to several temperature cycles with pure water before being used to measure conductance. The cell constant ($l/A\equiv{\kappa R}$ where $l/A$ is the distance to surface area ratio, $R$ is the resistance, and $\kappa$ is the specific conductance) was calculated as $2.959\pm0.050 \mathrm{m}^{-1}$ by measuring the resistance of different NaCl (\textit{aq}) solutions at 20 $^\circ$C. The specific conductance data of these NaCl (\textit{aq}) solutions were calculated from the correlation proposed by Peyman et al. \cite{peyman2007complex} The cell constant was re-measured after many temperature cycles and did not show a significant change.
\subsection{Measurement procedure}
Before performing a measurement, distilled water was used to clean the continuous flow unit for 3 hours at 523 K and 100 bar. After the reactor was cooled, distilled water was flowed for additional 3 hours to remove residual salts or oxidation residues. The desalination unit was then dried at 398 K overnight to remove any remaining water. After the system was cooled to 293 K, the installed electrode and titanium wires were connected to a frequency response analyzer (Solartron 1260A, Ameteck, Inc.). Before supplying the salt solutions, an \textit{Open} circuit measurement was always performed to correct for instrumental artifacts (measurement artifacts are discussed below in Sec. \ref{sec2.5:bulk resistance estimation}). Prior to performing a measurement, feed solution was continuously supplied to the reactor for one hour to remove any additional impurities and/or remaining water. After the effluent concentration stabilized, the feed flow rate was reduced to 2 mL/min. To examine whether the cell constant changed, the impedance was measured at room temperature, and then the pressure was increased to the operating pressure (250 bar). The vessel was then heated to the desired temperature. After the operating temperature reached the specified measurement temperature, impedance measurements were performed by holding the input voltage constant at $V=50 \mathrm{mV}$ over a frequency range of $10^0$ to $10^6$ Hz. To verify that the oxidation impurity level did not have a significant effect on the measurement results, the specific conductance of the effluent from the flow unit was measured \textit{ex-situ} by the handheld conductometer and the sodium selective electrode at 293 K. Up to the highest measurement temperature explored in this work (623 K), no considerable change ($<2 \%$) was observed when compared to the feed. Table 1 summarizes the measurement conditions covered in this work.
\subsection{Bulk resistance estimation}
\label{sec2.5:bulk resistance estimation}
The impedance spectra obtained from a frequency response analyzer are frequently affected by a myriad of measurement artifacts. These mainly arise from electrical leakage at the electrical feedthrough, inductance between long wires connected to electrodes, parasitic capacitance, and the internal circuit impedance of the electronic instruments \cite{edwards1997experimental,veal2014understanding}. Although ceramic insulators and sealants were installed in the system, the artifacts could not be removed completely. Among them, the wire series impedance (\textit{Short}) and the \textit{Open} circuit shunt impedance contribution from the apparatus were significant. 

\begin{figure}
	\begin{center}
	\includegraphics[width=0.45\textwidth]{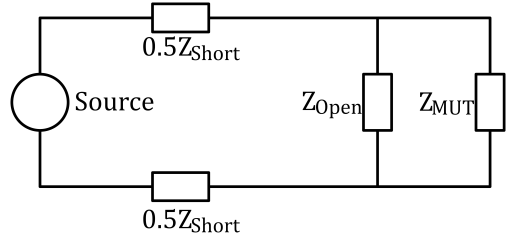}
	\caption{The equivalent circuit model for the \textit{in-situ} EIS measurement in the continuous flow cell. When the frequency response analyzer (Source) exerts an alternating current (AC) electric signal, the signal passes through the wire (half of the \textit{Short}) and is split into \textit{Open} and the material under test (MUT). The split ratio depends on the nature of the MUT. The electrical signal then passes through the other wire (the other half of the \textit{Short}) and is received by the analyzer. Each of the components could be modeled by analyzing the impedance spectra. The equivalent circuit models for all components are explained in the text.}
	\label{fig3: equivalent-circuit}
	\end{center}
\end{figure}

To subtract the contribution from these artifacts, an \textit{Open/Short} correction algorithm was used based on an equivalent circuit model proposed by Edwards et al. \cite{edwards1997experimental} Edwards et al. suggested that the \textit{Open} contribution could be approximated as an $p(R,C)$ (parallel) circuit, whereas the \textit{Short} could be modeled as a $s(R,L)$ (series) circuit. When the resistance of the working fluid is quite low (close to \textit{Short} conditions), the electrical leakage is insignificant and could be ignored. On the contrary, when the resistance of the working fluid is high enough to be comparable to the parasitic effect (e.g., pure water), the electrical leakage can have a significant effect on the measured spectra. According to the hypothesized equivalent circuit (Figure \ref{fig3: equivalent-circuit}), the impedance spectra of the material under test (MUT) $Z_\mathrm{MUT}$ is formulated as:
\begin{equation}
	Z_\mathrm{meas}(f)=Z_\mathrm{s}(f)+\left[Z_\mathrm{o}^{-1}(f)+Z_\mathrm{MUT}^{-1}(f)\right]^{-1}
	\label{eq:eq-circuit}
\end{equation}
where $Z_\mathrm{meas}$ is the measured impedance, and $Z_\mathrm{s}$ and $Z_\mathrm{o}$ are measured impedances of the \textit{Open} and \textit{Short} circuits.
To measure \textit{Open} and \textit{Short}, the following protocol was followed. First, the \textit{Short} impedance data were measured by inserting a copper wire between the electrodes at different temperatures from 293 to 723 K. The resultant spectra were well-fit by the following short circuit model
\begin{equation}
	Z_\mathrm{s}(f)=R_\mathrm{s}+(2\pi{i}f)^\alpha L_\mathrm{s}
\end{equation}
where $i$ is the imaginary unit number ($i^2=-1$), $f$ is the frequency, $R_\mathrm{s}$ is the resistance in $\Omega$, $L_\mathrm{s}$ is the inductance in H, and $\alpha$ is an exponent to account for non-ideality.
\onecolumn
\begin{longtable}{ccccc}
	\caption{Experimental conditions, specific conductance and equivalent conductance data obtained in NaCl (aq) solutions at different molal concentrations ($m$ [mol/kg]) along an isobar (250 bar). The relative deviation $\Delta\Lambda/\Lambda$ between the measurement results and the Fuoss-Hsia-Fern{\'a}ndez-Prini (FHFP) model was calculated as $(\Lambda-\Lambda_\mathrm{FHFP})/\Lambda_\mathrm{FHFP}$ multiplied by a factor of 100.}	\\
	\hline
	Temperature [K] & $\kappa$ [$\mathrm{{\mu}S/cm}$] & $\Lambda$ [$\mathrm{S\ cm^2/mol}$] & $\Lambda_\mathrm{FHFP}$ [$\mathrm{S\ cm^2/mol}$] & $\Delta\Lambda/\Lambda$ [\%] \\
	\hline
	\endfirsthead
	\hline Temperature [K] & $\kappa$ [$\mathrm{{\mu}S/cm}$] & $\Lambda$ [$\mathrm{S\ cm^2/mol}$] & $\Lambda_\mathrm{FHFP}$ [$\mathrm{S\ cm^2/mol}$] & $\Delta\Lambda/\Lambda$ [\%] \\
	\hline
	\endhead
	\hline \multicolumn{5}{r}{Continued}
	\endfoot	
	\hline	\endlastfoot
	\multicolumn{5}{c}{$m=1.697\ $mmol/kg} \\
	373	&	554.07	&	336.82	&	339.41	&	0.76	\\
	398	&	650.05	&	402.84	&	418.85	&	3.82	\\
	423	&	755.88	&	478.92	&	497.09	&	3.65	\\
	448	&	859.38	&	558.44	&	572.6	&	2.47	\\
	473	&	960.52	&	642.38	&	644.67	&	0.36	\\
	498	&	1027.8	&	710.4	&	713.32	&	0.41	\\
	523	&	1101.5	&	790.86	&	779.27	&	-1.49	\\
	548	&	1149.1	&	862.92	&	843.91	&	-2.25	\\
	573	&	1177.9	&	934.32	&	909.42	&	-2.74	\\
	598	&	1193.1	&	1015.5	&	979.39	&	-3.69	\\
	623	&	1173	&	1105.2	&	1061	&	-4.17	\\
	\multicolumn{5}{c}{$m=4.286\ $mmol/kg} \\
	373	&	1366.7	&	328.9	&	329.12	&	0.07	\\
	398	&	1624.5	&	398.55	&	406.89	&	2.05	\\
	423	&	1872.6	&	469.71	&	483.34	&	2.82	\\
	448	&	2115.9	&	544.33	&	556.94	&	2.26	\\
	473	&	2359. 8 	&	624.77	&	626.97	&	0.35	\\
	498	&	2534.8	&	693.54	&	693.42	&	-0.02	\\
	523	&	2665.7	&	757.67	&	756.96	&	-0.09	\\
	548	&	2776.3	&	825.33	&	818.81	&	-0.8	\\
	573	&	2808. 7 	&	881.88	&	880.92	&	-0.11	\\
	598	&	2815.2	&	948.43	&	946.28	&	-0.23	\\
	623	&	2744.1	&	1023.2	&	1020.3	&	-0.28	\\
	\multicolumn{5}{c}{$m=8.704\ $mmol/kg} \\
	373	&	2631.8	&	311.87	&	316.79	&	1.55	\\
	398	&	3195.9	&	386.07	&	392.83	&	1.72	\\
	423	&	3718.5	&	459.27	&	467.47	&	1.75	\\
	448	&	4136.6	&	523.98	&	539.19	&	2.82	\\
	473	&	4626	&	603.05	&	607.24	&	0.69	\\
	498	&	4899	&	659.93	&	671.6	&	1.74	\\
	523	&	5169.7	&	723.43	&	732.85	&	1.29	\\
	548	&	5361	&	784.61	&	792.12	&	0.95	\\
	573	&	5465.8	&	844.86	&	851.11	&	0.73	\\
	598	&	5447.3	&	903.33	&	912.27	&	0.98	\\
	623	&	5211.9	&	956.38	&	979.26	&	2.34	\\
	\multicolumn{5}{c}{$m=13.06\ $mmol/kg} \\
	373	&	3829.6	&	302.5	&	307.45	&	1.61	\\
	398	&	4593.8	&	369.89	&	382.29	&	3.24	\\
	423	&	5388.5	&	443.6	&	455.71	&	2.66	\\
	448	&	6061.6	&	511.76	&	526.18	&	2.74	\\
	473	&	6638.2	&	576.77	&	592.96	&	2.73	\\
	498	&	7034.1	&	631.57	&	655.99	&	3.72	\\
	523	&	7455.9	&	695.36	&	715.81	&	2.86	\\
	548	&	7675.2	&	748.6	&	773.47	&	3.22	\\
	573	&	7844.2	&	807.98	&	830.49	&	2.71	\\
	598	&	7774.3	&	859.02	&	888.9	&	3.36	\\
	623	&	7497.9	&	916.55	&	950.91	&	3.61	\\
	\multicolumn{5}{c}{$m=18.29\ $mmol/kg} \\
	373	&	4983.42	&	281.1	&	298.26	&	5.75	\\
	398	&	6083.4	&	349.8	&	371.96	&	5.96	\\
	423	&	7063.5	&	415.25	&	444.26	&	6.53	\\
	448	&	7967.8	&	480.37	&	513.61	&	6.47	\\
	473	&	8718.3	&	540.92	&	579.26	&	6.62	\\
	498	&	9440.2	&	605.24	&	641.11	&	5.6	\\
	523	&	9994	&	665.52	&	699.68	&	4.88	\\
	548	&	10416	&	725.33	&	755.9	&	4.04	\\
	573	&	10675	&	785	&	811.11	&	3.22	\\
	598	&	10549	&	832.04	&	866.83	&	4.01	\\
	623	&	10097	&	880.81	&	923.44	&	4.62	\\	
\end{longtable}
Note that instead of the normal inductor used by Edwards \cite{edwards1997experimental}, a modified inductor model was used to fit the slightly right-skewed characteristics of the observed \textit{Short} \cite{diard2012handbook}. No significant temperature dependence was found for the inductance parameters $L_\mathrm{1}$ and $\alpha$. They were obtained as $L_\mathrm{1}=1.086\times10^{-6}\pm3.3\times10^{-8}\ \mathrm{H}$ and $\alpha=0.98\pm0.007$. The resistance part showed approximately 10 \% increase when the temperature increased up to 723 K. Provided that the major contribution of the resistance comes from the titanium wire and the welded parts, this measurement result could be utilized to determine the temperature dependence of the \textit{Short}. However, since the introduction of copper wire leads to an additional contact resistance between the copper wire and the platinum electrode, the resistance (real part) cannot be derived reliably. Thus, instead of estimating the \textit{Short} contribution, the \textit{Short} model (a series of the resistor and the modified inductor) was included in the equivalent circuit model fitted to the impedance spectra. 

The \textit{Open} impedance was measured by adding no substance into the unit in the same temperature range used to measure the \textit{Short}. Both magnitudes and angles of the \textit{Open} impedance did not show a significant temperature dependence but were more complicated than a simple RC parallel circuit. The response could be approximated as $s(p(R_\mathrm{o1},Q_\mathrm{o1}),Q_\mathrm{o2})$ where $Q$ stands for the constant phase element ($Q=q_1(2\pi if)^\mathrm{q_2}$; $q_1>0$ and $0\leq{q_2}\leq1$). However, we decided to use the \textit{Open} impedance spectra directly rather than relying on a model, since including \textit{Open} model parameters in the Markov Chain Monte Carlo (MCMC) method associated with the models makes numerical convergence difficult within a short run.

After obtaining the impedance spectra of the MUT, three methods to estimate the specific conductance were evaluated. The first method was the Zimmerman technique, which basically exploits the power-law dependence of the real part of the impedance at low frequency. It is given as:
\begin{equation}
	Z_\mathrm{r}(\omega)=R_\mathrm{MUT}+b(2\pi f)^\mathrm{c};\ R_\mathrm{MUT}=\frac{1}{\kappa}\frac{l}{A}
\end{equation}
where $Z_\mathrm{r}$ is the real part of the complex spectra, and $R_\mathrm{MUT}$, $b$ and $c$ are adjustable parameters \cite{zimmerman2012limiting}. The obtained $R_\mathrm{MUT}$ can be further used to estimate the specific conductance $\kappa$ by using the cell constant $l/A$.

Another analysis technique is to use the phase angle. Typically, an aqueous electrolyte system changes from capacitive ($\theta\equiv\arctan(Z_\mathrm{i}/Z_\mathrm{r})<0$) to resistive ($\theta=0$) as the measurement frequency decreases beyond a certain value. As the frequency decreases further, the contribution of the electrical double layer becomes dominant and the system again shows a capacitive behavior ($\theta<0$). The phase angle $\theta$ has its maximum (or close to zero) at the frequency where the system becomes resistive. To find the phase angle minimum, a spline function was constructed for $|Z|=\sqrt{Z_\mathrm{r}^2+Z_\mathrm{i}^2}$ and $\theta$ as a function of the frequency, respectively. The resistance $R_\mathrm{MUT}$ was set to be $|Z|(f)$ whose frequency matches with the condition that the phase angle is equal to its maximum. 

The other method examined involves establishing and fitting an equivalent circuit model to the entire spectrum \cite{balashov2009experimental}. Several different models of this class have been suggested for electrolyte solutions \cite{lima2017electric}. We used an equivalent circuit model $s(p(R_\mathrm{b},C_\mathrm{b}),p(R_\mathrm{G},Q_\mathrm{G}),Q_\mathrm{EDL})$, which gives
\begin{equation}
	Z_\mathrm{MUT}=\frac{1}{R_\mathrm{b}^{-1}+C_\mathrm{b}^{-1}}+\frac{1}{R_\mathrm{G}^{-1}+Q_\mathrm{G}^{-1}}+Q_\mathrm{EDL}
\end{equation}
In this model, $R_\mathrm{b}$ and $C_\mathrm{b}$ are the bulk resistance and capacitance, $Q_\mathrm{G}$ and $R_\mathrm{G}$ are the components for representing the grain boundary contribution, and $Q_\mathrm{EDL}$ is the electrical double layer contribution modeled as the constant phase element.\cite{ishai2013electrode} The measured \textit{Open} was inserted to Eq. \ref{eq:eq-circuit} when the equivalent circuit-based regression analysis was performed.

The Markov Chain Monte Carlo (MCMC) algorithm implemented in our previous work was employed to fit the equivalent circuit models \cite{yoon2020dielectric}. The MCMC method based on the Metropolis-Hastings algorithm is simple, free from parameter initialization in principle, and accurate \cite{kawahara2018unique}. In this algorithm, the model parameters ($R,\ C,\ L, \cdots$) are first initialized by examining the arc shape and the low-frequency part of the impedance spectra \cite{lvovich2012impedance}. These initial parameters are used to calculate the first model spectrum. Then, one of the model parameters is randomly chosen and changed from $p$ to $p(1+\Delta)$ where $\Delta$ is an arbitrary number between zero and a rate constraint. A global rate constraint or individual rate constraints can be determined by evaluating the convergence behavior of the Markov chain. A new model spectrum is calculated based on the changed parameters. Then, the performance of the parameter sets before and after the change is compared based on the residual function, which is defined as:
\begin{equation}
	\chi^2=\sum_{\mathrm{i=1}}^{\mathrm{N_\mathrm{p}}}||Z_\mathrm{i}^\mathrm{model}-Z_\mathrm{i}^\mathrm{exp}||^2
\end{equation}
where $N_\mathrm{p}$ is the number of measurement points, $Z_\mathrm{i}^\mathrm{model}$ and $Z_\mathrm{i}^\mathrm{exp}$ are the impedance data at the $i^\mathrm{th}$ frequency from the model and the experiment, and $||n||$ indicates the modulus (magnitude) of a complex number $n$. When $f=\exp(-\chi_\mathrm{after}^2/4+\chi_\mathrm{before}^2/4)$ is higher than unity, the changed parameter is accepted. Otherwise, a random number between zero and unity is generated. If the random number is lower than $f$, the new parameter set is accepted although it increases the discrepancy between the model and the experimental result. This procedure makes it possible to escape from the local minima in parameter space. If the random number is higher than $f$, the change is rejected, and the original parameter set is used in the next calculation step. This MCMC procedure is repeated until no significant change is observed in the parameter set. 

When the initial parameters were properly chosen by examining the impedance spectra \cite{lvovich2012impedance}, the Markov chain typically showed convergence within the burn-in run consisting of 5,000 steps. Hence, we use the burn-in run of 5,000 steps and explored the parameter space during a production run of 10,000 steps. After the MCMC run, the parameter set which showed the minimum $\chi^2$ was chosen as an optimal parameter set.

\subsection{Data analysis}
Most previous studies \cite{zimmerman1995new,zimmerman2012limiting,bannard1975effect,gruszkiewicz1997conductance,sharygin2001tests,sharygin2002multiple,corti2008electrical,ho1994electrical,pearson1963electrical,quist1968electrical,zimmerman2007new,noyes1903electrical,fogo1954electrical,corwin1960conductivity} report the electrical properties not as specific conductance ($\kappa$) but as the equivalent conductance ($\Lambda$). The equivalent conductance is defined as $\Lambda\equiv{\kappa}/{N}$ where $N$ is the normality in a NaCl (\textit{aq}) solution. Since the cation charge is $+1$, the normality is same to the molarity in NaCl (\textit{aq}) solutions. The molality ($m$) of a solution can be converted into molarity ($c$) by the following equation.
\begin{equation}
	c=\frac{m\rho_\mathrm{s}}{mM+1000}
\end{equation}
where $m$ is the molarity, $M$ is the molar mass, $\rho_\mathrm{s}$ is the solution density [$\mathrm{kg/m^3}$]. In earlier works, $\rho_\mathrm{s}$ could be approximated as the density of pure water at the same temperature and pressure, which works well in dilute solutions \cite{ho1994electrical,quist1968electrical} but fails when the solution concentration is high. This work used the Helgeson-Kirkham-Flowers (HKF) model \cite{tanger1988calculation,shock1988calculation,shock1992calculation} in conjunction with the IAPWS-95 Equation of States (EoS) \cite{wagner2002iapws} following the procedure used by Zimmerman et al. \cite{zimmerman2012limiting} The HKF model estimates the solution volume ($V_\mathrm{s}, \mathrm{cm^3/g}$) as
\begin{equation}
	V_\mathrm{s}=\frac{1000}{\rho_\mathrm{w}}+\sum_\mathrm{i}m_\mathrm{i}V_\mathrm{i}^0+1000A_\mathrm{v}\frac{I\ln(1+bI^{1/2})}{b}
\end{equation}
where $\rho_\mathrm{w}$ is the density of water in kg/m\textsuperscript{3}, $V_\mathrm{i}^0$ is the standard partial molar volume of the species ($i=\mathrm{Na^+,\ Cl^-,}$ and $\mathrm{NaCl}$), $A_\mathrm{v}$ is the Debye-Hückel limiting slope for the apparent molar volume \cite{fernandez1997formulation}, $I$ is the ionic strength, and $b$ is a constant ($b=1.2\ \mathrm{(kg\cdot mol)}^{1/2}$). The static dielectric constant required in the HKF model was estimated based on the Fern{\'a}ndez model \cite{fernandez1997formulation}. The ionic radii data were calculated based on the suggestion by Shock et al. \cite{shock1992calculation}

There is a certain degree of inconsistency among literature data, even when the temperature and pressure (or density) conditions were similar (or equal) to each other \cite{zimmerman2012limiting}. Instead of comparing the measurement results directly, we compared our data with those generated from a conductance model. Among several models, we used the Fuoss-Hsia-Fern\'andez-Prini (FHFP) equation \cite{fernandez1969conductance} in conjunction with the correlations for the limiting equivalent conductance and the thermodynamic association constant recently proposed by Zimmerman et al. \cite{zimmerman2012limiting} To calculate the thermodynamic association constant and the limiting equivalent conductance, Zimmerman proposed the following expressions based on a weighted regression analysis on the collected literature data. 
\numparts
\begin{eqnarray}
	\log_\mathrm{10}K_\mathrm{a,m}=21.09+\frac{825.0}{T}-7.52\log_\mathrm{10}\rho_\mathrm{w}\\
	\log_\mathrm{10}\Lambda^0=1.673+\left(-0.939+\frac{22.76}{\rho_\mathrm{w}}\right)\log_\mathrm{10}\eta
\end{eqnarray}
\endnumparts
where $K_\mathrm{a,m}$ is the association constant in molality scale, $\Lambda^0$ is the limiting equivalent conductance [$\mathrm{S\ cm^2/mol}$], and $\eta$ is the dynamic viscosity of pure water in Poise. The dynamic viscosity was estimated based on the IAPWS 2008 formulation \cite{huber2009new}. 

The FHFP equation without ion association is expressed as:
\begin{equation}
	\Lambda=\Lambda_0-Sc^{1/2}+Ec\ln{c}+J_1c-J_2c^{3/2}
\end{equation}
where $\Lambda_0$ is the limiting equivalent conductance, $S$ is the Onsager limiting slope (a function of the limiting equivalent conductance, relaxation coefficient, and electrophoretic coefficient). $E$, $J_1$, and $J_2$ are the functions of molarity, reciprocal radius of the ionic atmosphere, the distance of closest approach, and the limiting conductance. A discussion about the FHFP equation and the corresponding coefficients can be found in the work by Fern\'andez-Prini and others \cite{fernandez1969conductance,linert1982analyzing,miyoshi1973comparison}.
When ion aggregation is considered, the FHFP equation becomes
\begin{eqnarray}
	\Lambda=\alpha&\left[\Lambda_0-S(\alpha c)^{1/2}+Ec\ln({\alpha c})+J_1(\alpha c)\right.\\\nonumber&\left.-J_2(\alpha c)^{3/2}\right]
	\label{eq:FHFP-pair}
\end{eqnarray}
In Eq. \ref{eq:FHFP-pair}, $\alpha$ is the degree of dissociation. It is calculated from the thermodynamic association constant $K_\mathrm{a,c}$ in molarity scale, which is given as:
\begin{equation}
	K_\mathrm{a,c}=\frac{1-\alpha}{\alpha^2(c/c^\mathrm{0})\gamma_\mathrm{c,\pm}^2}
\end{equation}
Here, $c^0$ is a hypothetical reference state ($1$ mol/L), and $\gamma_\mathrm{c,\pm}$ is the mean activity coefficient in molarity scale. It is calculated as:
\begin{equation}
	\ln\gamma_\mathrm{c,\pm}^2=\frac{-\kappa_\mathrm{dh}q_\mathrm{B}\alpha^{1/2}}{1+\kappa_\mathrm{dh}q_\mathrm{B}\alpha^{1/2}}
\end{equation}
We set the distance of the closest approach $q_\mathrm{B}$ to be equal to the Bjerrum radius, following the suggestion by Justice \cite{justice1971interpretation}. The calculated $\Lambda_\mathrm{FHFP}$ data were compared to our experimental results as well as some selected previous works in Sec. \ref{sec:results}.
\subsection{Computational Fluid Dynamics}
Since the measurement is performed in a continuous flow unit, it is quite possible that there is a non-negligible contribution of the pressure fluctuation and flow pattern to the measurement results. To study how operating conditions can affect the results, we carried out a series of CFD simulations. The goal of the simulations was to establish the velocity profiles existing between the probes and to determine if those local flow velocities could hinder/alter the experimental results.
\subsubsection{Physical properties}
The physical properties of the sodium chloride solutions, including bulk density ($\rho$), enthalpy ($H$), dynamic viscosity ($\eta$), isobaric heat capacity ($c_\mathrm{p}$), and thermal conductivity ($\lambda$), were estimated based on the correlations proposed by earlier works. Specifically, density ($\rho$), enthalpy ($H$), and dynamic viscosity ($\eta$) were calculated based on the equivalent temperature method \cite{michaelides1986nbs,driesner2007system1,driesner2007system2}. In this method, the physicochemical properties in NaCl (\textit{aq}) solutions at the condition of temperature $T$, pressure $p$ and the fraction $x$ is estimated by defining an equivalent temperature $T_\mathrm{q}$($q=\rho,\ H,\ $and $\eta$) that satisfies the following equation.
\begin{equation}
	q_\mathrm{w}(T_\mathrm{q},p)=q_\mathrm{b}(p,T,x)
	\label{eq:property-estimation}
\end{equation}
where $q_\mathrm{w}$ and $q_\mathrm{b}$ are the properties in pure water and that in the brine, respectively. The pure water properties were calculated from the Wagner-Pru{\ss} EoS \cite{wagner2002iapws} and the IAPWS 2008 formulation for the dynamic viscosity \cite{huber2009new}. The equivalent temperature $T_\mathrm{q}$ for each property was defined as $T_\mathrm{q}=a+bT$ where the coefficients $a$ and $b$ were functions of $T$ and $x$ (See Driesner et al. \cite{driesner2007system1,driesner2007system2} and Klyukin et al. \cite{klyukin2017revised} for the numerical details). Then, the isothermal compressibility ($\kappa_\mathrm{T}$) and heat capacity ($c_\mathrm{p}$) were obtained from the density and enthalpy data by performing a numerical differentiation. The thermal conductivity ($\lambda$) was calculated based on the correlation proposed by Wang and Anderko \cite{wang2012modeling}. 
\begin{figure}[!ht]
	\begin{center}
	\includegraphics[width=0.45\textwidth]{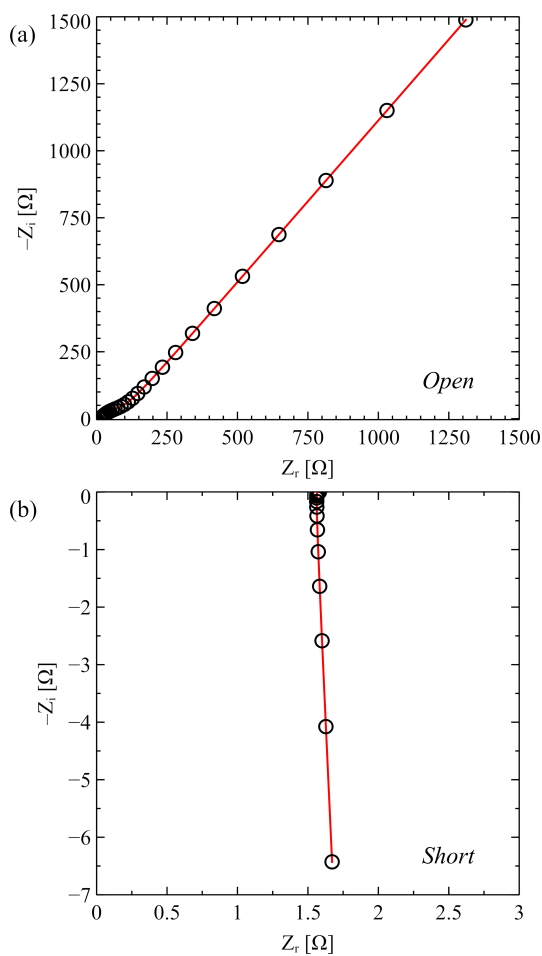}
	\caption{Nyquist plots of (a) \textit{Open} and (b) \textit{Short}. Measured spectra (black circles) could be well represented by the equivalent circuit model (red lines) proposed in this work. Note that the high-frequency part of the \textit{Short} has a positive phase angle, which suggests that the inductance plays an important role in determining the high-frequency behavior. \textit{Open} shows a rather complex behavior compared to what Edwards et al. reported whereas \textit{Short} can be modeled successfully with a series connection of the resistance and the modified inductor. When an ordinary inductor was used, the high-frequency arc part in the Nyquist plot of samples was slightly distorted but did not have a significant effect on the estimation of the bulk resistance.}
	\label{fig4: open-short}
	\end{center}
\end{figure}
\subsubsection{CFD simulation details}
\begin{figure*}[!ht]
	\includegraphics[width=\textwidth]{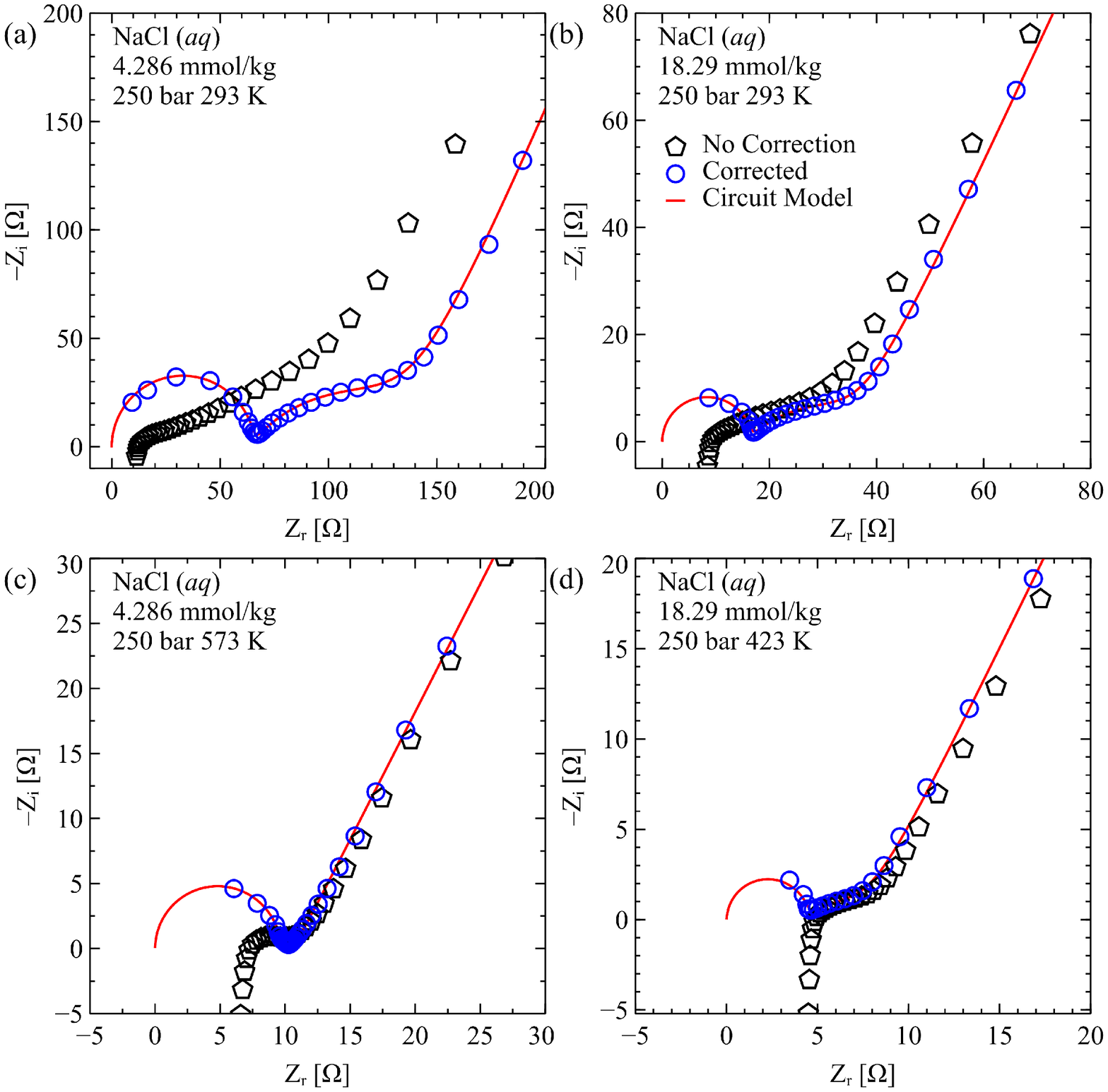}
	\caption{Nyquist plots of the impedance spectra in NaCl (aq) solutions at different conditions. (a) $m=4.29\ \mathrm{mmol/kg}$ and $T=293$ K, (b) $m=18.29\ \mathrm{mmol/kg}$ and $T=293$ K, (c) $m=4.29\ \mathrm{mmol/kg}$ and $T=573$ K, and (d) $c=18.29\ \mathrm{mmol/kg}$ and $T=423$ K. Black pentagons indicate raw impedance spectra, and blue circles denote the corrected one. The impedance spectrum cannot be used to estimate the bulk resistance in a reliable manner when no correction was done. After the correction, the spectrum shows a typical behavior observed in aqueous electrolytes. It consists of (1) the bulk response, (2) grain boundary contribution, and (3) the electrical double layer contribution.}
	\label{fig5: nyquist-plots}
\end{figure*}
A preliminary model of the electrode was constructed based on the experimental electrode geometry with slight modification. The two primary deviations between the simulation and experimental setup were (1) the titanium wires attached to the electrodes were simplified from cylindrical shapes to rectilinear shapes with the same hydraulic diameter and (2) the system outlet was defined to be 20 internal reactor diameters away from the probe locations. Due to the local geometric complexity of the reactor design, a hybrid meshing approach was applied with hexahedral cells used throughout the domain, but with a fluid region near the probe and wires. The fluid region near probes utilized four-sided tetrahedral cells (that were later converted to polyhedral elements for the internal volume) and five-sided wedge cells for each boundary region. A total of 3,257,462 cells were used in the initial CFD analyses with a minimum of 9 inflation layers surrounding each surface. The magnified mesh surrounding the probes highlights the emphasis on the inflation layers applied to resolve the local boundary layers. Aspect ratios were maintained below a value of 100 with the average cell aspect ratio being 7.61 with a standard deviation of 13.57.

Flow rates within the reactor were limited to the range of 2 mL/min to 4 mL/min at the temperature range from 323 to 623 K. Taking into consideration the low flow rates, a laminar viscous model was applied. Fitted curves were applied for density, thermal conductivity, and specific heat to capture the nonlinearity of pure water material being investigated, which is in line with Eq. \ref{eq:property-estimation} and properties presented in Table 1. All simulations were conducted at an operating pressure of 250 bar. A pseudo-transient, coupled solver was applied to predict pressure and velocity values within the flow field.

The inlet was treated as a mass flow inlet with a constant mass flow value based upon the 2 to 4 mL/min volumetric flow range and the density at the incoming 298 K inlet temperature. The outlet was placed approximately $20D$ above the probes, ensuring that the exit boundary would not affect the local boundary layer present between and around the probes. The outer wall of the reactor was treated as a constant temperature boundary condition based on the prescribed 323 to 623 K operation temperature range. A total of 32 separate grid-to-grid (GGI) interfaces were applied between the interior solid elements (probes, separator, and wiring) and the fluid body. The interfaces were treated as coupled boundaries with heat transfer allowed through them. Solid elements have a minimum of four cells spanning the thickness in order to capture any conduction present through the solid. The solution was initialized at 298 K and the mean velocity in the vertical direction based upon the mass flow prescribed at the inlet. Relative residuals for momentum and continuity equations were solved to a value of $10^{-4}$. The second order discretization schemes were applied to each variable.
\section{Results and Discussion}
\label{sec:results}
\subsection{Algorithm validation}
\begin{figure}[!ht]
	\begin{center}
	\includegraphics[width=0.45\textwidth]{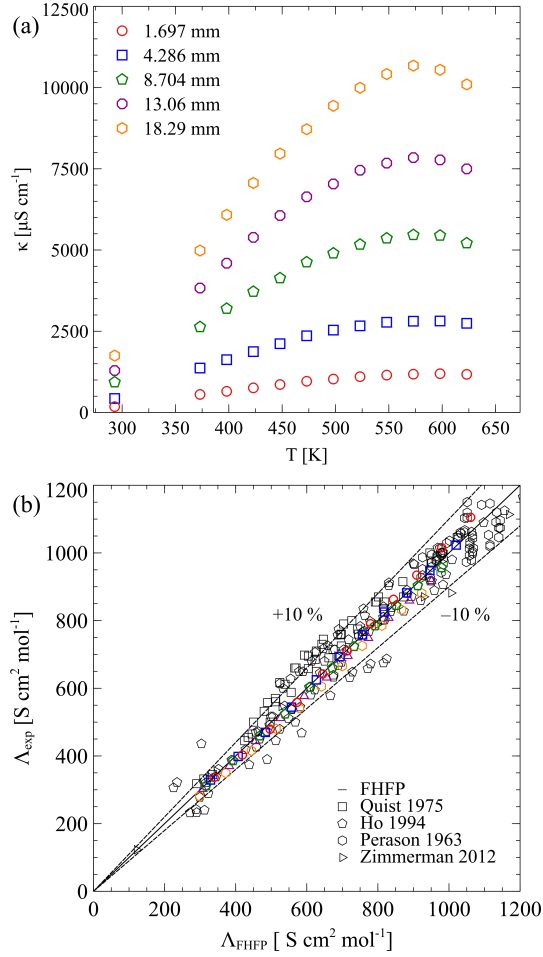}
	\caption{(a) Specific conductance data in NaCl (\textit{aq}) solutions from 298 to 623 K at 250 bar. The temperature where $\kappa$ shows its maximum decreased as the salt concentration increased. This result suggests that the ion pair formation in NaCl (\textit{aq}) solutions occurs at lower temperature when the concentration increases. (b) Parity plot between the experimental measurements from this work and literature and the FHFP model. Our experimental data (colored points) are consistent with the FHFP model proposed by Zimmerman et al. with an absolute average relative deviation (AARD) of 2.5 \%. The dashed lines denote 10 \% deviation.}
	\label{fig6: conductance-data}
	\end{center}
\end{figure}
We first examine the \textit{Open} and \textit{Short} measurement results and the impact of the algorithmic correction procedure. 

Figure \ref{fig4: open-short} shows Nyquist plots of the \textit{Open} and \textit{Short}. The equivalent circuit models could be well-fitted to the measured spectra. We also tested if a simpler \textit{Short} model used by Edwards et al. \cite{edwards1997experimental} that consists of only the resistor and the inductor can be applied. In this case, the high-frequency arc that corresponds to bulk response was slightly distorted. Figure 5 shows the influence of the \textit{Open/Short} correction algorithm on the impedance spectra obtained at different conditions. In low concentration NaCl (\textit{aq}) solutions (0.01 \textit{wt}\%), an elimination of the \textit{Open} contribution results in the significant resistance increase. On the other hand, an accurate determination of the \textit{Short} contribution becomes important when the specific conductance of the material under test becomes low (i.e., in high-temperature and high-concentration NaCl (\textit{aq}) solutions). 

In all solutions, the usual and expected behavior of aqueous electrolytes was restored when the \textit{Open/Short} correction was performed. The complex resistance spectrum shows a high-frequency arc on the Nyquist plot (complex plane), which represents the bulk resistance and capacitance. In the low-frequency region, a positive slope is observed, which represents the contribution of the electrical double layer. In the mid-frequency domain, there is another element that should be considered in addition to the bulk behavior and the electrode double layer. This behavior is frequently observed when the grain boundaries within an electrode has a significant impact on the impedance spectra \cite{gerstl2011separation}.

Although the equivalent circuit models could be fitted to the measured spectra after the \textit{Open/Short} correction, it should be noted that some parameters, especially the bulk capacitance term, have a high uncertainty when the salt concentration is high. This mainly arises from the fact that the high loss MUTs have a high RC characteristic frequency. The RC characteristic frequency $f_\mathrm{RC}$ is defined as the frequency where the absolute magnitude of the bulk resistance and the reactance become equal. It is given as:
\begin{equation}
	f_\mathrm{RC}=\frac{\kappa}{2\pi\epsilon_\mathrm{r}\epsilon_\mathrm{0}}
\end{equation}
When the characteristic frequency $f_\mathrm{RC}$ is much higher than the measurement frequency, there are few data points that can be utilized to estimate the material's capacitance. In seawater at ambient conditions [NaCl (\textit{aq}) 3.5 \textit{wt}\%] for instance, $f_\mathrm{RC}$ becomes $\sim0.8\ \mathrm{GHz}$, which is much higher than the typical frequency limit of most frequency response analyzers. 

Next, we examined three methods to determine the bulk resistance of an MUT. Since the grain boundary contribution was not negligible in our experimental setup, it was necessary to control the fitting frequencies when utilizing the Zimmerman method. A blind use of the Zimmerman method typically yielded a higher resistance compared to the other two methods when the grain boundary contribution was significant (e.g., Figure \ref{fig5: nyquist-plots} (a)). By adjusting the fitting frequency range, the average bulk resistance values obtained from the Zimmerman method agreed well with the others (typically within 3 \%). However, we also observed that the bulk resistance obtained from the Zimmerman method fluctuated up to 40 \%, depending on the fitting frequency range. The phase angle-based method worked well in the low concentration salt solutions. The bulk resistance difference between the phase angle-based method and the equivalent circuit model was less than 1 \%. They started to deviate as the concentration increased, since the grain boundary contribution becomes comparable to that of the bulk resistance. In lossy materials, the impedance spectra became almost resistive ($Z_\mathrm{i}\sim0$) in the transition frequency range between the bulk contribution and the grain boundary contribution. In these samples, the resistance data obtained from the phase-angle based method was generally higher than that from the equivalent circuit method. Considering all these factors, we used the equivalent circuit method to determine the bulk resistance.
\begin{figure}[b]
	\begin{center}
	\includegraphics[width=0.45\textwidth]{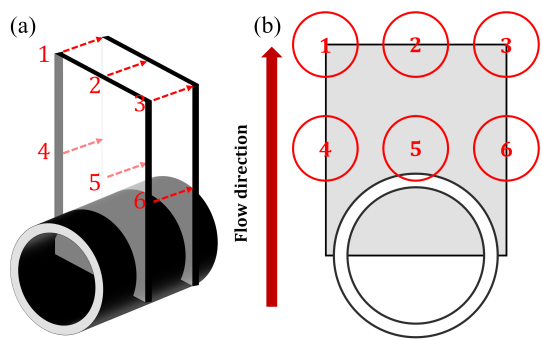}
	\caption{(a) Side view and (b) front view of the computational fluid dynamics (CFD) simulation paths. The flow direction in the apparatus was upwards. The CFD simulation paths for examining the velocity profile and pressure distribution are denoted in red.}
	\label{fig7: cfd-geometry}
	\end{center}
\end{figure}
\subsection{Measurement validation}
\begin{figure*}[!ht]
	\includegraphics[width=\textwidth]{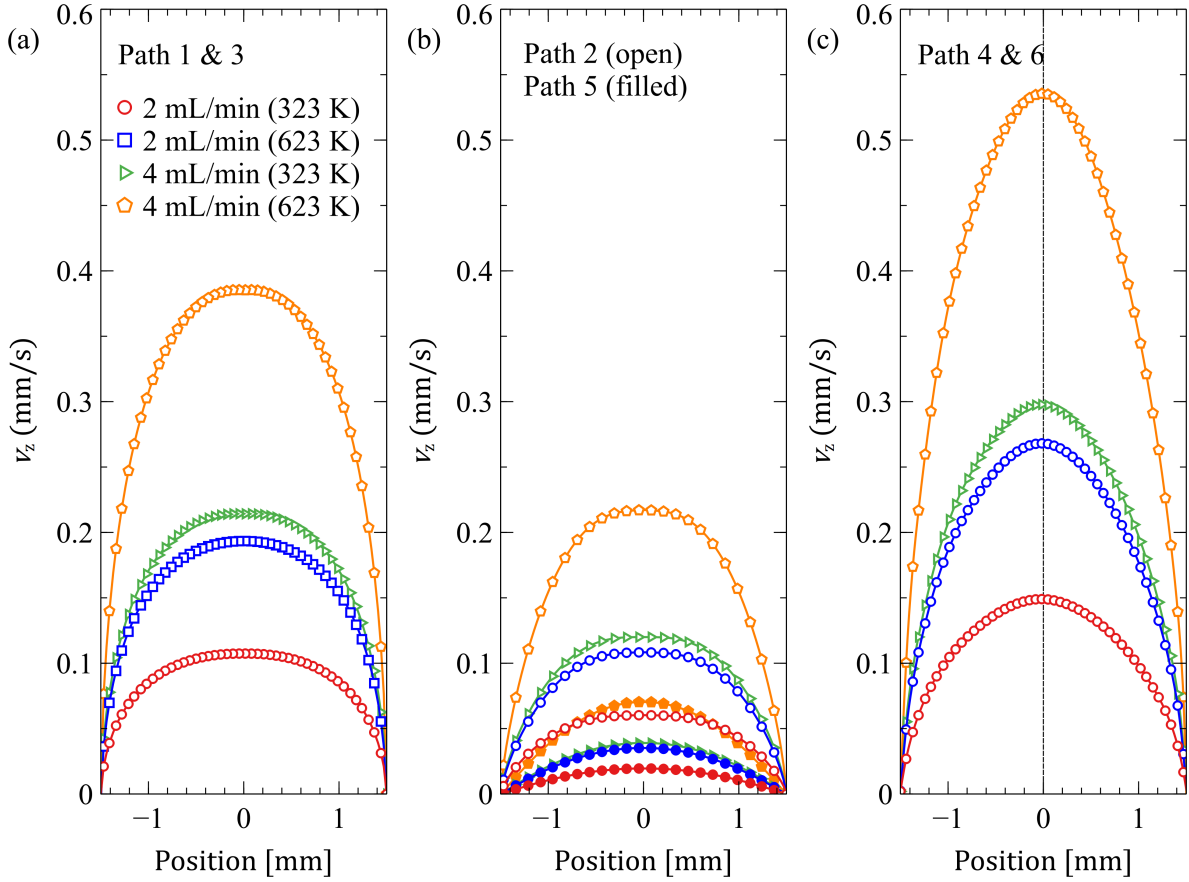}
	\caption{Velocity profiles between the two-parallel electrodes. The maximum velocity at the center between the parallel plates varies from 0.0703 mm/s to 0.536 mm/s, depending on the position. In (c), the velocity profiles are slightly asymmetric due to the presence of the aluminum oxide support.}
	\label{fig8: velocity-profile}
\end{figure*}
Figure \ref{fig6: conductance-data} (a) show the specific conductance ($\kappa$) in NaCl (\textit{aq}). Over the concentration and temperature ranges considered, the specific conductance data shows a monotonic increase below the temperature where $\kappa$ has its maximum. As the feed concentration increased, the conductance maximum temperature ($T_\mathrm{max}$) was shifted from 598 to 548 K, which agrees well with previous works \cite{bannard1975effect,quist1968electrical,zimmerman2012limiting}. Our previous MD simulation results demonstrated that the occurrence of the conductance maximum indicates the formation of neutral ion pairs \cite{yoon2019electrical,patel2021nacl}. This result suggests that the \textit{in-situ} implementation of the conductometric device is helpful for understanding the high-temperature behavior of ions in solutions. In measuring the specific conductance, the uncertainty (fluctuation level) was largest in the highest concentration (18.29 mmol/kg) solutions. Since we used an electronic scale, a sodium selective electrode, and a commercial handheld electrode, the solution preparation error is expected to be negligible. The uncertainty introduced by trace amounts of pre-existing salt in the distilled water should be negligible, considering its conductivity contribution ($<1\ \mathrm{{\mu}S/cm}$). The first main contributor of the data uncertainty was likely due to the the grain boundary effects. As the concentration increased, the grain boundary impedance $Z_\mathrm{G}$ became comparable to that of the bulk impedance ($Z_\mathrm{b}$). A careful choice of initial parameters was required to discern the grain boundary contribution and the bulk contribution. The relative magnitude of the fluctuation (uncertainty) obtained by iterating MCMC procedures with different initial parameters was up to 5 \% in the highest concentration solution. In the lowest concentrations, it was always below 0.5 \%. The second largest uncertainty contribution was likely from the $Z_\mathrm{s}$ (\textit{Short} impedance). The \textit{Short} impedance estimated from the MCMC procedure could fluctuate up to 5 \%. When the \textit{Short} resistance $R_\mathrm{s}$ was restricted to be below the values obtained in the copper wire measurement, the fluctuation level diminished below 1 \%.

Figure \ref{fig6: conductance-data} (b) compares the measurement results obtained in this work, the literature data, and the FHFP model calculation. The equivalent conductance data obtained from the model shows a reasonable agreement with the literature data. The parity line $\Lambda_\mathrm{FHFP}=\Lambda_\mathrm{exp}$ lies between the literature data by Quist and Marshall [denoted as Quist 1975 in Figure \ref{fig6: conductance-data} (b)] \cite{quist1968electrical} and Ho, Palmer and Mesmer [Ho 1994 in Figure \ref{fig6: conductance-data} (b)] \cite{ho1994electrical}. The experimental results by Zimmerman, Arcis and Tremaine \cite{zimmerman2012limiting} are closer to Quist and Marshall. Although the experimental results by Ho, Palmer and Mesmer are more scattered than Quist and Marshall, it should be noted that most data points agree with the FHFP model within 15 \%. The saturation data by Pearson and Benson \cite{pearson1963electrical} are scattered around the parity line. The equivalent conductance data measured in this work also agrees well with the FHFP model calculation results. The absolute average relative deviation (AARD) from the FHFP model is approximately 2.5 \%. This result suggests that the implemented device and methodologies in this work can be utilized for both scientific and engineering purposes. 

\subsection{CFD predictions}
The probes were relatively invariant to temperature within the simulation volume due to both the proximity of the constant temperature boundary condition and the distance from the inlet to the probes. Six paths were selected between the probes (Figure \ref{fig7: cfd-geometry}). Since the distance between the inner thermocouple tip and the electrode is close, solutions could serve as isothermal predictions of the flow, dependent upon fluid properties resulting from the temperature applied to the reactor wall. Three paths were placed at the top of the probe furthest away from the separator and three were positioned in the vertical center of the probes. The interior of the cylindrical separator served as the stagnation point for the fluid. 

Figure \ref{fig8: velocity-profile} shows the velocity profiles for the bounding test cases based on incoming flow and the applied wall temperature. The flow profiles between Path 4 and 6 and between Path 1 and 3 were symmetrical in behavior. Path 5 had the lowest flow peak flow, at less than 0.1 mm/s, due to its proximity to the separator. Path 4 and Path 6 both had peak flows of 0.536 mm/s for inlet volumetric flow of 4 mL/min and 623 K. Due to the orientation of the separator, the velocity at Paths 4 and 6 show a slightly asymmetric profile; the flow is accelerated to either side of the geometry due to the slotted aluminum oxide support. However, the influence of the aluminum oxide support was negligible. Near the top of the probes, at Paths 1 through 3, the flow has dispersed into a more uniform distribution and the effects of the aluminum oxide holder (support) are less pronounced despite boundary layers still existing around the probes themselves. Overall, the results suggest that the velocity distribution should not have a significant effect on the measurement results.
\begin{figure}
	\begin{center}
	\includegraphics[width=0.45\textwidth]{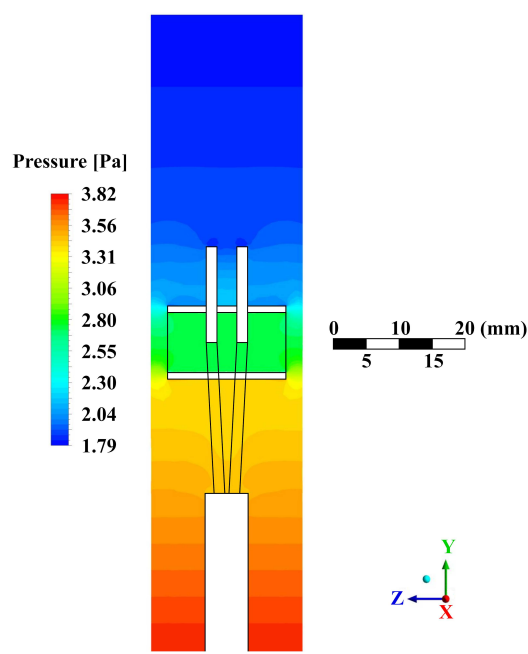}
	\caption{Pressure drop around the in-situ probe at 623 K and 250 bar. The volumetric flowrate of the feed solution was set to be 4 mL/min. The pressure drop across the electrode was only 1.9 Pa, which is negligible compared to the operating pressure.}
	\label{fig9: pressure-drop}
	\end{center}
\end{figure}

The local pressure drop distribution around the electrode was also examined (Figure \ref{fig9: pressure-drop}). At the highest velocity condition examined in this work (4 mL/min), the pressure drop over the probe, including the electrodes and the aluminum oxide support, was only approximately 1.9 Pa. Thus, the pressure drop between the electrode was not significant.
\section{Conclusions}
This work described the implementation details and a set of methodologies that can be used to obtain the specific conductance of flowing fluids \textit{in-situ} at high temperatures and pressures. Such measurements are applicable to monitoring thermal desalination processes or high-pressure extraction processes and can also be used to obtain the equivalent conductance, which facilitates understanding aqueous electrolyte behavior at elevated temperatures and pressures. In terms of hardware, a platinized platinum electrode was successfully inserted into a high-pressure unit through an electrical feedthrough (protected by active cooling) and electrical insulators were installed to remove safety hazards. Software methods included use of Markov Chain Monte Carlo together with an \textit{Open/Short} correction scheme, which was shown to effectively remove the measurement artifacts from raw impedance spectra. The corrected frequency-dependent resistance spectra showed behavior typically observed in frequency-dependent resistance spectra obtained \textit{ex-situ} without any artifacts. Three bulk resistance estimation methods were examined with the equivalent circuit-based method that includes the \textit{Open}, \textit{Short}, and the material contributions showing the smallest uncertainty compared to power-law or phase-angle based methods.

We also performed computational fluid dynamics (CFD) simulations to study possible effects of the sensor geometry on the velocity profile and the local pressure distribution around the electrode. By assuming an isothermal boundary condition, the CFD simulations demonstrate that the electrode geometry has a negligible impact on the pressure change, as well as the velocity profile, in the flowing aqueous electrolyte solutions.
 
In summary, this work documents the successful installation and experimental validation of a continuous flow cell equipped with an \textit{in-situ} conductometer, which shows a reasonable agreement with the literature data as well as the FHFP model calculation results. Based on the CFD predictions of less than 0.55 mm/s flow velocities between the reactor probes, it is not expected that experimental measurements were affected significantly by any of the higher-order and nonlinear flow/transport effects. In future work, variations in temperature and functionality of the reactor surface will be examined via additional calculations and numerical simulations to provide more detailed descriptions of regions within the reactor that are away from the probe locations. In addition, we will utilize the experimental capability described here to explore selective recovery of specific salts in either liquid-liquid extraction or desalination processes. 
\section*{Data availability statement}
All data supporting this work are included within the article.
\ack
This work was supported by the Director's Postdoctoral Fellow Program (20190653PRD4) and the Laboratory Directed Research and Development Program (20190057DR) at Los Alamos National Laboratory.
\section*{Conflict of interest}
The authors declare no known competing financial interests or personal relationships that could influence the work reported in this paper.
\section*{References}
\bibliography{main}
\end{document}